\documentclass[pra,showpacs,twocolumn,eqsecnum]{revtex4}

\usepackage{graphicx}
\usepackage[urlcolor=blue, hyperindex, colorlinks, bookmarks=true,linkcolor=black,citecolor=black]{hyperref}

\begin{document}

\newcommand{\sch}{Schr\"odinger }
\newcommand{\schs}{Schr\"odinger's }
\newcommand{\nn}{\nonumber}
\newcommand{\nl}{\nn \\ &&}
\newcommand{\dg}{^\dagger}
\newcommand{\bra}[1]{\langle{#1}|}
\newcommand{\ket}[1]{|{#1}\rangle}
\newcommand{\bl}{{\Bigl(}}
\newcommand{\br}{{\Bigr)}}
\newcommand{\erf}[1]{Eq.~(\ref{#1})}
\newcommand{\erfs}[2]{Eqs.~(\ref{#1}) and (\ref{#2})}
\newcommand{\erft}[2]{Eqs.~(\ref{#1}) -- (\ref{#2})}
\newcommand{\rt}[1]{\sqrt{#1}}
\newcommand{\Bra}[1]{\Big{\langle}{#1}\Big{|}}
\newcommand{\Ket}[1]{\Big{|}{#1}\Big{\rangle}}

\newcommand{\psizk}[1]{\ket{\psi_{z}({#1})}}
\newcommand{\psizb}[1]{\bra{\psi_{z}({#1})}}
\newcommand{\lpsizk}[1]{\ket{\tilde{\psi}_{z}({#1})}}
\newcommand{\lpsizb}[1]{\bra{\tilde{\psi}_{z}({#1})}}
\newcommand{\lpsixyk}[1]{\ket{\tilde{\psi}_{\bf X,Y}({#1})}}
\newcommand{\lpsixyb}[1]{\bra{\tilde{\psi}_{\bf X,Y}({#1})}}
\newcommand{\lpsixk}[1]{\ket{\tilde{\psi}_{\bf x}({#1})}}
\newcommand{\lpsixb}[1]{\bra{\tilde{\psi}_{\bf x}({#1})}}
\newcommand{\psiqk}[1]{\ket{\psi_{{\bf q}(t')}({#1})}}
\newcommand{\psiqb}[1]{\bra{\psi_{{\bf q}(t')}({#1})}}
\newcommand{\lpsiqk}[1]{\ket{\tilde{\psi}_{{\bf q}(t')}({#1})}}
\newcommand{\lpsiqb}[1]{\bra{\tilde{\psi}_{{\bf q}(t')}({#1})}}

\newcommand{\red}[1]{\color{red}{#1}\color{black}\ }
\newcommand{\blu}[1]{\color{blue}{#1}\color{black}\ }

\title{The interpretation of non-Markovian stochastic Schr\"odinger
equations as a hidden-variable theory}
\date{8 July 2003}
\author{Jay Gambetta}
\affiliation{Centre for Quantum Dynamics, School of Science,
Griffith University, Brisbane 4111, Australia}
\author{H. M. Wiseman} \email{h.wiseman@griffith.edu.au}
\affiliation{Centre for Quantum Dynamics, School of Science,
Griffith University, Brisbane 4111, Australia}

\begin{abstract}
Do diffusive non-Markovian  stochastic Schr\"odinger equations (SSEs) for open quantum systems have a physical interpretation? In a recent paper [Phys. Rev. A {\bf 66}, 012108 (2002)] we investigated this question using the orthodox interpretation of quantum mechanics. We found that the solution of a non-Markovian SSE represents the state the system would be in at that time if a measurement was performed on the environment at that time, and yielded a particular result. However, the linking of solutions at different times to make a trajectory is, we concluded, a fiction. In this paper we investigate this question using the modal (hidden variable) interpretation of quantum mechanics. We find that the noise function $z(t)$ appearing in the non-Markovian SSE can be interpreted as a hidden variable for the environment. That is, some chosen property (beable) of  the environment has a definite value $z(t)$ even in the absence of measurement on the environment. The non-Markovian SSE gives the evolution of  the state of the system ``conditioned'' on this environment hidden variable. We present the theory for diffusive non-Markovian SSEs that have as their Markovian limit SSEs corresponding to homodyne and heterodyne detection, as well as one which has no Markovian limit.
\end{abstract}


\maketitle

\section{Introduction}

In nature it is very unlikely to find a system existing in
isolation; usually it is immersed in an environment (or bath). In
quantum mechanics we label this type of system an open quantum
system \cite{Car93}. To determine the evolution we must solve the
\sch equation
\begin{equation}\label{sch}
  d_{t}\ket{\Psi(t)}=-\frac{i}{\hbar}\hat{H}_{\rm
  uni}(t)\ket{\Psi(t)},
\end{equation} where $\ket{\Psi(t)}$ and $\hat{H}_{\rm uni}(t)$ are the quantum state and Hamiltonian for
the complete universe (system and bath). That is, $\ket{\Psi(t)}$
belongs to the Hilbert space ${\cal H}_{\rm uni}={\cal H}_{\rm
sys}\otimes{\cal H}_{\rm bath}$. Due to the large Hilbert space of
the bath (${\cal H}_{\rm bath}$) it is convenient to describe the
system by its reduced state. This is defined as
\begin{equation}\label{ReducedState} \rho_{\rm red}(t)={\rm
Tr}_{\rm bath}[\ket{\Psi(t)}\bra{\Psi(t)}],
\end{equation} and operates only in the ${\cal H}_{\rm sys}$.

It has been shown \cite{Nak58,Zwa60} by a projection-operator
method that we can write a general master equation for the reduced
state as
\begin{equation} \label{Master} d_{t}{\rho}_{\rm
red}(t)=-\frac{i}{\hbar}[\hat{H}(t),\rho_{\rm
red}(t)]+\int_{0}^{t}\hat{\cal K}(t,s)\rho_{\rm red}(s) ds,
\end{equation} where $\hat{H}(t)$ operates only in ${\cal H}_{\rm sys}$
and $\hat{\cal K}(t,s)$ is the ``memory time'' superoperator. It
operates on $\rho_{\rm red}(s)$ and represents how the bath has
affected the system in the past. The problem with this equation is
that in general $\hat{\cal K}(t,s)$ cannot be explicitly
evaluated.

In recent papers
\cite{DioStr97,DioGisStr98,StrDioGis99,GasNag99,Cre00,Bud00,GamWis02,
GamWis03,BasGhi02,Bas02} non-Markovian stochastic \sch equations
(SSEs) have been proposed which allow an alternative procedure for
solving the reduced state. A non-Markovian SSE is a stochastic
equation for the system state $\psizk{t}$, conditioned on some
noise function $z(t)$. We consider only continuous SSEs, although
discontinuous ones have also been proposed \cite{JacCol00}. The
SSE has the property that when the projector for $\psizk{t}$ is
averaged over all the possible $z(t)$ one obtains $\rho_{\rm
red}$(t). That is,
\begin{equation}\label{Ensemble}
\rho_{\rm red}(t)={\rm E}[\psizk{t}\psizb{t}],
\end{equation}
where E[...] denotes an ensemble average over all possible
$z(t)$'s.

 When using non-Markovian SSEs to solve the reduced state it
turns out that in general we cannot explicitly evaluate
$\psizk{t}$. However, as shown in Refs.
\cite{YuDioGisStr99,GamWis02b} we can use perturbative techniques
to find approximate solutions. Here we are interested not in how
to solve the non-Markovian SSEs, but in how to {\em interpret}
them.

When Markovian SSEs (the Markovian limit of non-Markovian SSEs)
were introduced in quantum optics their first interpretation was
as a numerical tool \cite{DalCasMol92}. Another interpretation was
that Markovian SSEs represent {\em objective} (that is,
independent of any observer)
 trajectories for the system \cite{Dio88,Gis89,Pea89}.
In this interpretation the SSE is seen as a (stochastic and
nonlinear)
 modification of the \sch equation, modelling  how
state reduction (collapse of the wave function) occurs in open
quantum systems. However, in recent times it has been generally
accepted that Markovian SSEs are evolution equations for the
system state conditioned on continuous measurement of the bath
\cite{Car93,WisMil93a,WisMil93,Wis96b}.

A Markovian bath is one for which the bath correlation time is
much less than the decoherence time of the system. For such a
system one can envisage making repeated measurements on the bath
on a time scale  infinitesimal compared to the system decoherence
time but large compared to the bath correlation time. Thus these
measurements do not lead to any back-action of the system from the
bath. But the measurement of the bath does yield information about
the system. This can be described as the action of a measurement
operator (as appears in generalized  quantum measurement theory
\cite{QM}) for an infinitesimal time interval \cite{Wis96b}, and
keeps the system in a pure state. It can be reexpressed as a SSE
in which the stochastic variable $z(t)$ is related to the results
of the continuous measurements.
 The stochastic trajectory of this conditioned system state
generated by the Markovian SSE is often referred to as a quantum
trajectory \cite{Car93}. Different detection procedures
(measurements on the bath) result in different unravelings. A few
common examples are direct detection \cite{Car93}, homodyne
\cite{Car93,WisMil93a}, and heterodyne detection \cite{WisMil93}.

In the light of the quantum trajectory interpretation of Markovian SSEs,
we return to the interpretation of non-Markovian SSEs.
Obviously one interpretation is that they are simply a numerical
tool used to generate the reduced state. However, after the
success in finding a physical interpretation for Markovian SSEs, it is
natural to seek something beyond this trivial interpretation. Moreover, we
have previously shown \cite{GamWis02} that there are different non-Markovian
unravelings, and that these correspond to different measurement
schemes (homodyne and heterodyne) in the Markovian limit. Thus it
is natural to seek an interpretation of non-Markovian SSEs beyond
that of being just numerical tools \cite{Walter}.

In Ref. \cite{GamWis02} we came to the conclusion that under the
orthodox interpretation \cite{QM} of quantum mechanics, the
solution of a non-Markovian SSE at time $t$ is the state the
system will be in, if at that time a measurement was performed on
the bath and yielded a result $z$. Thus the non-Markovian SSE
under this view has no interpretation; it is just a stencil used
to calculate the conditioned system state at a particular time
$t$. In other words, the linking of $\psizk{t}$ (or $z(t)$) with
itself at times less then $t$ turns out to be a convenient
fiction.

Unlike the Markovian case, it is not possible to derive the SSE by
 continuously measuring the bath because a
 non-Markovian bath has a non-negligible correlation time.
Thus if a measurement is made at time $t$, collapsing the bath
state at that time, this will affect the state of the bath
interacting with the system in the future. That is, the
measurement on the bath will cause a backaction on the system and
hence the average evolution for a system state conditioned on
continuous measurement of the bath will not be that of
\erf{ReducedState}. In fact, a continuous measurement on a bath
with a nonzero correlation time will lead to a quantum Zeno
effect, radically altering the average evolution of the system.
If, on the other hand, the measurement on the bath is not done
continuously,
 then the system state will not remain in a pure state as the system and bath
 will become entangled in the time between measurements on the bath. Thus it
seems safe to conclude that there is no continuous measurement
interpretation, and in fact we are forced to accept that SSEs are
only a numerical tool which could be used to determine the system
state at a particular time conditioned on a particular measurement
result $z$.

Since  orthodox quantum mechanics fails to give a satisfying
interpretation for non-Markovian SSEs, in this paper we turn to a
{\em non}orthodox approach: the modal interpretation of quantum
mechanics
\cite{Bel84,Bub97,VerDie95,BacDic99,Sud00,SpeSip01b,GamWis03b}.
This interpretation, unlike the orthodox interpretation, has as
its basic goal to keep reality intact. That is, the values of some
observables (the hidden variables) really exist before we measure
them. Because the observables have an objective reality from now
on we will refer to them as properties or beables (after Bell
\cite{Bel84}). Just as in the orthodox theory, where it is
impossible to simultaneously measure all observables, in the modal
theory it is impossible to give all observables property status
The best-known example of such an interpretation is Bohmian
mechanics for particles \cite{Boh52} in which position is the
preferred observable (property).

We expect a modal interpretation to be applicable to non-Markovian
SSEs because we can use it to assign definite properties to the
bath, as occurs in the orthodox theory when the bath is measured,
{\em without} invoking such a measurement. In this way we avoid
the backaction problem which arose in the orthodox theory. While
the bath is ascribed definite properties, the system is described
as a purely quantum system. But, because of the entanglement
between the system and the bath, we can define a system state
associated with (or ``conditioned'' on) a particular value for the
bath property. If the bath properties are described by rank-one
projective measures on ${\cal H}_{\rm bath}$, then the conditioned
system state will be {\em pure}. Averaging over the conditioned
system state would reproduce the non-Markovian reduced state
matrix $\rho_{\rm red}(t)$, just as in \erf{Ensemble}.

Since the value of the bath hidden variables change in time, the
conditioned system state will evolve in time also, and this time
there is a meaningful relation between the conditioned system
state at different times. A particular set of bath properties
amounts to a particular {\em decomposition} (of the unit operator
on ${\cal H}_{\rm bath}$), and leads to a particular {\em
unraveling} of the non-Markovian master equation (\ref{Master}).
We will use these terms interchangeably.

In this paper we show that for a suitable choice of bath
properties we  can reproduce all of the non-Markovian SSEs
discussed above, as well as one which has not been previously
considered. The noise function $z(t)$ appearing in the
non-Markovian SSE is simply a function of the values of the bath
hidden variables. The system state $\psizk{t}$ is the system state
conditioned on the bath properties having the values giving
$z(t)$. Moreover, the system state guides the values of the hidden
variables of the bath.

The format of this paper is as follows: In Sec. \ref{Sec2} a
summary
 of the modal interpretation of quantum mechanics is
presented.
 In Sec. \ref{Sec3} we give the microscopic model for
 non-Markovian dynamics for an open quantum system that underlies
 all of our work.  In Sec. \ref{Sec4} we show how modal dynamics can be applied
 to such models, and derive
 the  hidden variable interpretation of non-Markovian SSEs  for three different unravelings.
 These correspond to
 objective values for the position of each bath harmonic oscillators (as in
 the Bohmian interpretation), for the quadrature of the collective bath field, and for
 the coherent amplitude of the collective bath field. The first of these
 has no Markovian limit, while the latter two have homodyne and heterodyne
 detection as their Markovian limit.  Lastly in Sec. \ref{Sec5} we conclude with
 a discussion and directions for future work.

\section{Modal Interpretation of Quantum Mechanics} \label{Sec2}

In this section we give a brief overview of the modal
interpretation of quantum mechanics; for a more detailed
description see Refs.
\cite{Bel84,Bub97,VerDie95,BacDic99,Sud00,SpeSip01b,GamWis03b}.
The basic idea of this view of quantum mechanics is that certain
observables have a definite reality independent of measurement,
whereas in the orthodox interpretation the measurement defines the
reality of the observable. To explain this we consider measurement
of the observable $Z$.  The notation we use to represent an
observable is
\begin{equation}\label{Obser}
   {Z}=\{(z_{n},\hat{\pi}_{n})\}.
\end{equation}  That is, ${Z}$ is represented by a set of
paired elements. Each pair represents the value $z_{n}$ and the
corresponding projector $\hat{\pi}_{n}$. The projectors are
orthogonal and
 form a decomposition of unity:
\begin{equation}
 \sum_n \hat\pi_n = \hat{1}
\end{equation}
  In \erf{Obser} for simplicity we have
only considered the case where ${Z}$ is time independent.
We can in this case also define an operator $\hat{Z}$ which is equivalent
to \erf{Obser}, by
\begin{equation}\label{Z}
 \hat{Z}=\sum_{n}z_n\hat\pi_n.
\end{equation}

In the orthodox interpretation $Z$ has a definite value $z_n$ if
and only if $\ket{\Psi(t)}=\ket{\Psi_{n}(t)}$ (an eigenstate of
$\hat{Z}$). However, in general
$\ket{\Psi(t)}=\sum_{n}c_{n}(t)\ket{\Psi_{n}(t)}$, which implies
that $Z$ has all values contained in the sum; it is not well
defined. Upon measurement, by the introduction of a reduction
equation [$\ket{\Psi(t)}\rightarrow\ket{\Psi_{n}(t)}$] the value
of $Z$ becomes $z_n$ with probability $|c_{n}|^{2}$.

In the modal interpretation we choose one projective measure,
$\{\hat\pi_n\}$, as the preferred measure. This then determines
which observables can be given property status. To explain why the
property takes the value $z_n$ we introduce an extra quantum
state, the {\em property state},
\begin{equation}\label{propertystate}
\ket{\Psi_{n}(t)}=\hat\pi_{n}\ket{\Psi(t)}/\sqrt{N},
\end{equation} where $N$ is a normalization constant. This state propagates in time
along with $\ket{\Psi(t)}$, except it is stochastic in nature
(jumps between different $n$).
 It is interpreted as the
actual state of the universe, by the eigenstate eigenvalue it has
a definite value ($z_n$) for $Z$. The stochastic dynamics (rates
at which it jumps between different $n$) is determined by
$\ket{\Psi(t)}$ and so in this interpretation $\ket{\Psi(t)}$ is
called the guiding state.

For the purposes of this paper \erf{Z} is actually not general
enough. In Ref. \cite{GamWis03b} we showed that this theory can be
extended to positive operator measures (POMs) \cite{Hel76}, that
is
\begin{equation} \label{genproperty}
  Z=\{(z_{n}, \hat{F}_{n})\},
\end{equation} where
$\hat{F}_{n}$ is an effect (or POM element) with
$\sum_{n}\hat{F}_{n}=\hat{1}$. In \erf{genproperty}, $z_{n}$ is
the value of the effect, which could be a real number, a complex
number, a string of numbers, or even a statement (yes/no).

In Ref. \cite{GamWis03b} we showed that by implementing Naimark's
theorem, modal dynamics can be extended to include POMs. Naimark's
theorem says that if we enlarge the Hilbert space of the universe
from ${\cal H}_{\rm uni}$ to ${\cal K}={\cal H}_{\rm
uni}\otimes{\cal H}_{\rm aux}$, we can define a projector
$\hat{\Pi}_{n}$ in ${\cal K}$ such that
\begin{equation} \label{Nai}
\bra{\Psi(t)}\hat{F}_{n}\ket{\Psi(t)}=\bra{\Psi(t)}\bra{\phi}\hat{\Pi}_{n}\ket{\phi}\ket{\Psi(t)},
\end{equation} for all $\ket{\Psi(t)}\in{\cal H}_{\rm uni}$ and
for all possible $n$. $\ket{\phi}\bra{\phi}$ is called the Naimark
projection of ${\cal K}$ onto ${\cal H}_{\rm uni}$.  To work out
the set $\{\hat{\Pi}_{n}(t)\}$ it is necessary to introduce
another projector $\hat{\Pi}_{N+1}(t)$, such that
\begin{eqnarray}
\sum_{n}^{N+1}\hat{\Pi}_{n}(t)&=&\hat{1}_{\rm uni+ aux},
\end{eqnarray}
and
\begin{eqnarray}
\hat{\Pi}_{n}(t)\hat{\Pi}_{m}(t)&=&\hat{\Pi}_{n}(t)\delta_{nm}
\end{eqnarray} is satisfied for $n$, $m$,=$1,...,N+1$. The set of projectors in this enlarged
Hilbert space is called the Naimark extension of $\hat{F}_{n}(t)$
\cite{Hel76}. Worked examples of this are shown in \cite{Hel76}.
In this enlarged space the observable defined by \erf{genproperty}
becomes the property
\begin{equation}
  Z=\{(z_{n}, \hat{\Pi}_n)\},
\end{equation}
or equivalently
\begin{equation}
 \hat{Z}=\sum_{n}^{N+1}z_n\hat\Pi_n.
\end{equation} Here $\{\hat\Pi_n\}$ is the preferred projective
measure in ${\cal K}$. The guiding state becomes
\begin{equation} \label{PomState}
  \ket{\Phi(t)}=\ket{\Psi(t)}\otimes\ket{\phi},
\end{equation} where $\ket{\Psi(t)}$ is still the solution to the \sch
equation (\ref{sch}). The property state becomes
\begin{equation}\label{PomPropertystate}
  \ket{\Phi_{n}(t)}=\hat{\Pi}_{n}\ket{\Phi(t)}/\sqrt{ N}.
\end{equation}
That is, when describing POMs the universe is described by a
property state in the enlarged Hilbert space, which does not
factorize into a universe state and an auxiliary state. This can
be regarded merely as a mathematical construction to give reality
to POMs. Note that the value of $z_{N+1}$ is irrelevant as
$\hat{\Pi}_{N+1}$ projects into the null space of $\ket{\Phi(t)}$.
That is this value will occur with zero probability.

The modal dynamics [the stochastic evolution of the states in
\erf{propertystate} or \erf{PomPropertystate}] is found using the
method originally proposed by Bell \cite{Bel84} and generalized in
Refs. \cite{BacDic99,Sud00,GamWis03b} to include time dependent
projectors and POMs. Define $P_{n}(t)$ as the probability that the
system is in the $n^{\rm th}$ state at time $t$. Assuming a
Markovian process, by which we mean that the probability of being
in state $m$ at time $t+dt$ only depends on the state at time $t$,
we can write a master equation for $P_{n}(t)$ as
\begin{equation}\label{Master2}
  d_t P_{n}(t)=\sum_{m}[T_{nm}(t)P_{m}(t)-T_{mn}P_{n}(t)],
\end{equation} where $T_{nm}$ are
transition rates. For $n=m$, $T_{nn}$ (which is negative) is a
measure of the rate at which state $n$ losses probability.

Defining a probability current $J_{nm}(t)$ as
\begin{equation}\label{probcurrent}
J_{nm}(t)=T_{nm}(t)P_{m}(t)-T_{mn}P_{n}(t),
\end{equation} results in $J_{nm}(t)=-J_{mn}(t)$ and
allows us to rewrite the probability master equation as
\begin{equation} \label{master3}
d_t P_{n}(t)=\sum_{m}J_{nm}(t).
\end{equation} Given $J_{nm}(t)$ and $P_n(t)$, there
are many possible transition rates satisfying \erf{master3}. One
solution, chosen by Bell \cite{Bel84} is as follows.

For $J_{nm}(t)<0$,
\begin{eqnarray}\label{t1}
  T_{nm}(t)&=&0, \\
  T_{mn}(t)&=&-J_{nm}(t)/P_{n}(t),
\end{eqnarray} and for $J_{nm}(t)>0$
\begin{eqnarray}
  T_{nm}(t)&=&J_{nm}(t)/P_{m}(t),\\
  T_{mn}(t)&=&0. \label{t4}
  \end{eqnarray}
This is only one of the infinitely many solutions. These are found
by adding an extra term $T^{0}_{nm}(t)$ to $T_{nm}(t)$, where
$T^{0}_{nm}(t)$ is constrained only by
\begin{equation}
T_{nm}^{0}(t)P_{m}(t)-T_{mn}^{0}(t)P_{n}(t)=0.
\end{equation}

In Ref. \cite{GamWis03b} we showed that one possible solution for
$J_{nm}(t)$ is
\begin{eqnarray}
  J_{nm}(t)&=&2{\rm Im}\{\bra{\Phi(t)}\hat{\Pi}_{n}[\hat{H}_{\rm uni
}(t)\otimes\hat{1}_{\rm
aux}]\nl\times\hat{\Pi}_{m}\ket{\Phi(t)}\}{/\hbar}.
\end{eqnarray}   Note that this is only one of infinitely many possible
currents, as we can add any current $J_{nm}^{~0}(t)$ to
$J_{nm}(t)$ which satisfies $  \sum_{m} J_{nm}^{~0}=0,$
 to give a valid probability current. For the purposes of this paper we
only consider the simple solutions [not containing the extra
$T_{nm}^{0}(t)$ and $J_{nm}^{~0}(t)$ terms].

The above dynamics only describe a discrete decomposition.
For non-Markovian SSEs we must consider continuous decompositions.
 In Ref. \cite{GamWis03b} we showed that
provided the Hamiltonian is at most quadratic in the conjugate
variable to the chosen hidden variable, then the above dynamics
reduces to a deterministic theory. The only stochasticity is due
to the random initial conditions of the hidden variable. That is,
it is similar to Bohmian mechanics \cite{Boh52}. In fact if the
preferred projective measure is chosen to be the position
projective measure ($\hat{\pi}_{x}=\ket{x}\bra{x}$) then Bohmian
mechanics is the continuum limit.

We define continuous decompositions by the preferred projective
measure $\{\hat{\Pi}_{\{q_k\}}=\ket{\{q_{k}\}}\bra{\{q_{k}\}}\}$,
which in turn defines $\{\hat{q}_{k}\}$ as the operators for the
hidden variables.
 Here we have
introduced a notation for the set of hidden variables ($\{q_{k}\}$)
whose relevance will become important in the following sections.
 The decomposition of unity is
\begin{equation}
  \int  \ket{\{q_{k}\}}\bra{\{q_{k}\}} \prod_{k} dq_k = \hat 1,
\end{equation}
For this set of hidden variables we will see later that it is also
useful to define a single property $Z(t)$ as
\begin{equation} \label{genproperty2}
  Z(t)=\{(z(\{q_{k}\},t), \ket{\{q_{k}\}}\bra{\{q_{k}\}})\},
\end{equation} where $z(\{q_{k}\},t)$ is a function of the values
of the hidden variables $\{q_{k}\}$ and $t$.

As in Bohmian mechanics the evolution of each value of the hidden
hidden variable $q_{k}(t)$ (or the corresponding property state)
can be found by the differential equation
\begin{equation}\label{diff}
  d_{t}q_{k}(t)=v_{k}(\{q_{j}\},t)|_{q_{k}=q_{k}(t)},
\end{equation} where $v_{k}(\{q_{j}\},t)$ is the velocity field. This is defined
as \cite{GamWis03b}
\begin{equation} \label{velocityField}
  v_{k}(\{{q}_{j}\},t)=\frac{{\rm Re}[\bra{\Phi(t)}{\{q_{j}\}}\rangle\bra{\{q_{j}\}}\hat{v}_{k}(t)\ket{\Phi(t)}]}
  {\bra{\Phi(t)}{\{q_{j}\}}\rangle\bra{\{q_{j}\}}\Phi(t)\rangle}.
\end{equation} Here
 $\hat{v}_{k}(t)$ is the $k^{\rm th}$ component of the
velocity operator and is defined as
\begin{equation} \label{velocityO}
  \hat{v}_{k}(t)=-\frac{i}{\hbar}[\hat{q}_k,\hat{H}_{\rm uni}(t)\otimes\hat{1}_{\rm aux}].
\end{equation}

\section{Non-Markovian dynamics for open quantum systems} \label{Sec3}

The aim of this section is to outline the underlying dynamics used
to generate non-Markovian SSEs. Firstly we assume that there are
only two systems, a system of interest and a bath. That is, the
 Hamiltonian of the universe is
\begin{equation} \label{HamiltonianUni}
\hat{H}_{\rm sys}\otimes\hat{1}+\hat{1}\otimes\hat{H}_{\rm
bath}+\hat{V}.
\end{equation} The system Hamiltonian $\hat{H}_{\rm sys}$ is split into two terms, these being
$\hat{H}_{\Omega}$ and $\hat{H}$. The bath is modelled by a
collection of one-dimensional harmonic oscillators. In terms of
the bath annihilation and creation  operators, $\hat{a}_{k}$ and
$\hat{a}_{k}^{\dg}$, the Hamiltonian for the bath is
\begin{equation}\label{HamiltonianBath}
\hat{H}_{\rm
bath}=\sum_{k}\hbar\omega_{k}\hat{a}_{k}\dg\hat{a}_{k}.
\end{equation}

The interaction Hamiltonian $\hat{V}$ we assume is linear in the
bath amplitude, and has the form
\begin{equation}\label{HamiltonianInteraction}
\hat{V}=i\hbar\sum_{k}[\hat{L}g^{*}_{k}\hat{a}_{k}\dg-\hat{L}\dg
g_{k}\hat{a}_{k}],
\end{equation} where $\hat{L}$ is the system lowering operator and $g_{k}$ is the coupling strength of the $k^{\rm th}$
mode to the system. Without loss of generality we can take the
$g_{k}$'s to be real, absorbing any phases in the definitions of
the bath operators.

For calculation purposes we define an interaction frame such that
the fast dynamics placed on the state by the Hamiltonians
$\hat{H}_{\Omega}$ and $\hat{H}_{\rm bath}$ is moved to the
operators. The unitary evolution operator for this transformations
is
\begin{equation}\label{UnitaryFree}
\hat{U}_{0}(t,0)=e^{-i(\hat{H}_{\Omega}\otimes\hat{1}+\hat{1}
\otimes\hat{H}_{\rm bath})(t-0)/{\hbar}}.
\end{equation} Thus the combined state in the interaction frame is define as
\begin{equation}\label{UnitarySch} \ket{\Psi(t)}=\hat{U}_{0}\dg(t,0)\ket{\Psi(t)_{\rm Sch}},
\end{equation} and an arbitrary operator $\hat{A}$ becomes \begin{equation} \hat{A}_{\rm
int}(t)=\hat{U}_{0}\dg(t,0)\hat{A}\hat{U}_{0}(t,0). \end{equation} This allows
 us to write the \sch equation as
\begin{equation} \label{IntSchEquation} d_t\ket{{\Psi}(t)} =-\frac{i}{\hbar}\hat{H}_{\rm
uni}(t) \ket{\Psi(t)},
\end{equation} where $\hat{H}_{\rm
uni}(t) =\hat{H}_{\rm int}(t)+\hat{V}_{\rm int}(t)$. Here
$\hat{H}_{\rm int}(t)$ refers to $\hat{H}(t)$ in the interaction
picture and the form of the latter is
\begin{eqnarray} \label{HamiltonianInteractionInt} \hat{V}_{\rm
int}(t)= i\hbar\sum_{k}[\hat{L}g_{k}e^{i\Omega_{k}
t}\hat{a}_{k}\dg-\hat{L}\dg g_{k}e^{-i\Omega_{k} t}\hat{a}_{k} ],
\end{eqnarray} where $\Omega_{k}=\omega_{k}-\Omega$.  Here we have finally
restricted the form of $\hat{H}_{\Omega}$ to be such that $\hat{L}$ in the
interaction picture simply rotates in the complex plane at frequency $\Omega$.
 That is $\hat{L}_{\rm int}(t)=
\hat{L}e^{-i\Omega t}$.

\section{Non-Markovian stochastic \sch equations derived}\label{Sec4}

In this section we show that diffusive non-Markovian SSEs have an
interpretation under the modal interpretation of quantum
mechanics. To do this we choose a decomposition such that the
preferred projectors have the form
\begin{equation}\label{pro1}
\hat{\pi}_{\{q_k\}}=\ket{\{q_{k}\}}\bra{\{q_k\}}_{\rm bath}
\otimes\hat{1}_{\rm sys}.
\end{equation} This means the bath is given definite properties,
while the system is treated as a purely quantum system, which
nevertheless influences the bath values via the coupling
Hamiltonian. The different unravelings correspond to different
choices of $\ket{\{q_{k}\}}\bra{\{q_k\}}_{\rm bath}$. For the
overcomplete unravelings, like the coherent state unraveling
\cite{DioGisStr98,GamWis02} we have to use a POM \cite{GamWis03b}.
This means we have to use the enlarged Hilbert space preferred
projector $\hat{\Pi}_{\{q_k\}}$, which in general form is
\begin{equation}\label{pro2}
\hat{\Pi}_{\{q_k\}}=\ket{\{q_{k}\}}\bra{\{q_k\}}_{\rm bath +
aux}\otimes\hat{1}_{\rm sys}.
\end{equation}

With these projectors, the property states $\ket{\Psi_{n}(t)}$
[similarly for $\ket{\Phi_{n}(t)}$] can be factorized as
\begin{equation}\label{PomPropertystate2}
  \ket{\Psi_{n}(t)}=\ket{\{q_{k}\}}\ket{\psi_{\{q_{k}\}}(t)},
\end{equation} where $\ket{\psi_{\{q_{k}\}}(t)}$ is called the
conditioned system state.  It receives this name because it lives
entirely in ${\cal H}_{\rm sys}$ and is conditioned on the bath
values $\{q_{k}\}$. The form of $\ket{\psi_{\{q_{k}\}}(t)}$ is
\begin{equation}\label{conditionState}
\ket{\psi_{\{q_{k}\}}(t)}=\bra{\{q_k\}}\Psi(t)\rangle/\sqrt{ N},
\end{equation} where the normalization constant is defined as
\begin{equation}\label{N}
N=\langle\Psi(t)\ket{\{q_k\}} \bra{\{q_k\}}\Psi(t)\rangle.
\end{equation} This is the state of the system conditioned on the
bath hidden variables having values ${\{q_{k}\}}$.

For an actual trajectory (in the sense of Ref.~\cite{GamWis02}),
the bath values
  ${\{q_{k}(t)\}}$ are time dependent. This state
becomes $\ket{\psi_{\{q_{k}(t)\}}(t)}$ and represents the state of
the system conditioned on the bath having this trajectory. That
is, it is continuous in time and the differential equation of this
state will represent its evolution. In Refs. \cite{GamWis02} and
\cite{GamWis03} we showed that by starting with
\erf{conditionState}, the time derivative of this equation gives
diffusive non-Markovian SSEs. Thus in this paper we will not
reproduce these derivations, but instead show that by using our
velocity operator technique we can rederive the actual
trajectories for ${\{q_{k}}(t)\}$. This shows that diffusive
non-Markovian SSEs have a modal interpretation. If fact, because
the orthodox interpretation only gives an interpretation for the
solutions of non-Markovian SSE at a particular time (time of
measurement), we believe that the only nontrivial interpretation
of non-Markovian SSEs is a modal interpretation.

Before we consider specific unravelings we would like to note that
the velocity field, \erf{velocityField}, can be written in terms
of the conditioned system state as
\begin{equation} \label{velocityField2}
v_{k}(\{q_{k}\},t)={\rm
Re}[\bra{\psi_{\{q_{k}\}}(t)}\overrightarrow{\hat{
v}_{k}(\{q_{k}\},t)}\ket{\psi_{\{q_{k}\}}(t)}],
\end{equation} where
\begin{equation}
\overrightarrow{\hat{
v}_{k}(\{q_{k}\},t)}\ket{\psi_{\{q_{k}\}}(t)}\equiv{\bra{\{q_{k}\}}\hat{
v}_{k}(t)\ket{\Psi(t)}}/{\sqrt{N}}.
\end{equation}
This results in the following differential equation for the bath
values:
\begin{equation}\label{VelocityB}
d_{t}{q}_{k}(t)={\rm
Re}[\bra{\psi_{\{q_{k}(t)\}}(t)}\overrightarrow{\hat{
v}_{k}(\{q_{k}(t)\},t)}\ket{\psi_{\{q_{k}(t)\}}(t)}],
\end{equation} where
$\ket{\psi_{\{q_{k}(t)\}}(t)}=\ket{\psi_{\{q_{k}\}}(t)}|_{\{q_{k}=q_{k}(t)\}}$.

\subsection{Position unraveling} \label{position}

The first unraveling we consider is the position unraveling. This
results when we chose a preferred projective measure of the form
\begin{equation}\label{prox}
\{\hat{\pi}_{\{q_k\}}=\hat{\pi}_{\{x_k\}}=\ket{\{x_{k}\}}\bra{\{x_k\}}_{\rm
bath} \otimes\hat{1}_{\rm sys}\},
\end{equation} where $\ket{\{x_{k}\}}$ is the multimode eigenstate of the position
operators
$\hat{x}_{k}=(\hat{a}_{k}\dg+\hat{a}_{k})/\sqrt{2}$.

To simplify the overall equation we define a noise function $z(t)$
as
\begin{eqnarray}\label{z}
  z(t)&=&\sum_{k}g_{k}\sqrt{2}x_{k}(t)e^{-i\Omega_{k}t}.
\end{eqnarray}
In Ref. \cite{GamWis03} we showed that the non-Markovian SSE for
the position unraveling is
\begin{eqnarray}\label{SSE2}
d_{t}\ket{{\psi}_{z}(t)}&=&\Big{\{}-\frac{i}{\hbar}\hat{H}_{\rm
int}(t)+(\hat{L}-\langle\hat{L}\rangle_{t})z^{*}(t)-(\hat{L}\nl-\langle\hat{L}\rangle_{t})
[\hat{B}_{z}(t)+\hat{D}_{z}(t)]+\langle(\hat{L}-\langle\hat{L}\rangle_{t})[\hat{B}_{z}(t)\nl+\hat{D}_{z}(t)]\rangle_{t}
-(\hat{L}\dg-\langle\hat{L}\dg\rangle_{t})[\hat{A}_{z}(t)+\hat{C}_{z}(t)]\nl
+\langle(\hat{L}\dg-\langle\hat{L}\dg\rangle_{t})[\hat{A}_{z}(t)+\hat{C}_{z}(t)]\rangle_{t}
\nl\times\Big{\}} \ket{{\psi}_{z}(t)},
\end{eqnarray}
where $\ket{{\psi}_{z}(t)}\equiv\ket{{\psi}_{\{x_{k}(t)\}}(t)}$
and $\langle
\hat{L}\rangle_{t}=\bra{{\psi}_{z}(t)}\hat{L}\ket{{\psi}_{z}(t)}$.
The four operators $\hat{A}_{z}(t)$, $\hat{B}_{z}(t)$,
$\hat{C}_{z}(t)$ and $\hat{D}_{z}(t)$ are defined as ansatzen to
functional derivatives. It turns out that in general these
operators are not solvable. The perturbation techniques outlined
in Refs. \cite{YuDioGisStr99} and \cite{GamWis02b} can be applied
to this non-Markovian SSE to give a perturbative solution. Given
that there is no Markovian limit to this equation, however, it is
unclear whether such perturbative methods would be effective.

We also showed, after considerable effort, that
\begin{equation}\label{SDE}
  d_{t} x_{k}(t)={[\langle\hat{L}\rangle_{t}g_{k}
  e^{i\Omega_{k}t} +\langle\hat{L}\dg\rangle_{t} g_{k} e^{-i\Omega_{k}
  t}]}/{\sqrt{2}}.
\end{equation}
Integrating this gives
\begin{equation} \label{orthodx}
x_{k}(t)=x_{k}(0)+\int_{0}^{t} ds {[\langle\hat{L}\rangle_{s}g_{k}
  e^{i\Omega_{k}s} +\langle\hat{L}\dg\rangle_{s} g_{k} e^{-i\Omega_{k}
  s}]}/{\sqrt{2}}
\end{equation}
where $x_{k}(0)$ is the random variable chosen from the
distribution
\begin{equation}
  P(\{x_{k}\},0)=|\bra{\{x_{k}\}}\{0_{k}\}\rangle|^{2}
  =\prod_{k}\frac{\exp(-x_{k}^{2})}{\sqrt{\pi}}.
\end{equation}
That is, we have chosen the initial condition
$\ket{\Psi(t)}=\ket{\{0_{k}\}}\ket{\psi(0)}$.

To show that \erf{VelocityB} does give the same trajectories for
the values $\{x_{k}(t)\}$ as in \erf{orthodx}, we apply the
Hamiltonians defined in Sec. \ref{Sec3} to \erf{velocityO}, with
$q_{k}=x_{k}$. This gives
\begin{equation}
  \hat{v}_{k}(t)=[g_{k}e^{i\Omega_{k}t}\hat{L}+g_{k}e^{-i\Omega_{k}t}
\hat{L}\dg]/{\sqrt{2}},
\end{equation} as $[\hat{x}_{k},\hat{a}_{k}]=-1/\sqrt{2}$ and
$[\hat{x}_{k},\hat{a}_{k}\dg]=1/\sqrt{2}$. Substituting this into
\erf{velocityField2} gives a velocity field of the form
\begin{eqnarray}
  {v}_{k}(\{x_{k}\},t)&=&[g_{k}e^{i\Omega_{k}t}\bra{\psi_{\{x_{k}\}}(t)}
  \hat{L}\ket{\psi_{\{x_{k}\}}(t)}+g_{k}e^{-i\Omega_{k}t}\nl\times
\bra{\psi_{\{x_{k}\}}(t)}\hat{L}\dg\ket{\psi_{\{x_{k}\}}(t)}]/{\sqrt{2}}.
\end{eqnarray} Thus \erf{VelocityB} immediately reproduces \erf{SDE}, thereby confirming that the modal
theory does give the same non-Markovian SSEs, as found with the
orthodox theory.

\subsection{Quadrature unraveling}

The next unraveling we consider is what we call the quadrature
unraveling. In Ref. \cite{GamWis02} we show that this unraveling
only exists for certain environments, such that for every mode $k$
in the bath there exists another mode, which we can label $-k$,
such that $\Omega_{-k}=-\Omega_{k}$ and $g_{-k}=g_{k}$. These
simply imply that the modes coupled to the system come in
symmetric pairs about the system frequency $\Omega$. The form of
the preferred projective measure for this unraveling is
\begin{equation}\label{proq}
\{\hat{\pi}_{\{q_k\}}=\hat{\pi}_{\{X_{k}^{+},Y_{k}^{-}\}}=\ket{\{X_{k}^{+},Y_{k}^{-}\}}\bra{\{X_{k}^{+},Y_{k}^{-}\}}_{\rm
bath} \otimes\hat{1}_{\rm sys}\},
\end{equation}
where
$\ket{\{X_{k}^{+},Y_{k}^{-}\}}=\prod_{k>0}\ket{X_{k}^{+},Y_{k}^{-}}$
where $\ket{X_{k}^{+},Y_{k}^{-}}$ is the two mode entangled (EPR)
state
\begin{equation} \label{Qstatex}
\ket{X_{k}^{+},Y_{k}^{-}}=\int \frac{dx'}{\rt{2\pi}}
\Big{|}\frac{X_{k}^{+}-x'}{\rt{2}}\Big{\rangle}_{-k}\Big{|}\frac{X_{k}^{+}+x'}{\rt{2}
}\Big{\rangle}_{k}e^{iY_{k}^{-}x'}.
\end{equation} Here $\ket{(X_{k}^{+}+x')/\rt{2}}_{k}$ is an eigenstate of $\hat{x}_{k}$,
 and similarly $\ket{(X_{k}^{+}-x')/\rt{2}}_{-k}$ for $\hat{x}_{-k}$.
Equation (\ref{Qstatex}) is an eigenstate of both the operators
\begin{eqnarray}
\hat{X}_{k}^{+}&=&(\hat{x}_{k}+\hat{x}_{-k})/{\rt{2}}\label{X+}, \\
\hat{Y}_{k}^{-}&=&(\hat{y}_{k}-\hat{y}_{-k})/{\rt{2}}, \label{Y-}
\end{eqnarray}
 where $\hat{x}_{k}$ and $\hat{y}_{k}$ are the quadratures of
$\hat{a}_{k}$:
 \begin{equation} \label{aIintoxAndy}
\hat{a}_{k}=({\hat{x}_{k}+i \hat{y}_{k}})/{\rt{2}}.
\end{equation}
As in the position unraveling we define a noise function $z(t)$ as
\begin{equation} \label{QuadratureNoiseFunction}
z(t)=\sum_{k>0}{2} g_{k} [ X_{k}^{+}(t)
\cos(\Omega_{k}t)+Y_{k}^{-}(t)\sin(\Omega_{k}t)],
\end{equation}
which by definition is real. In Ref. \cite{GamWis02} we showed
that the non-Markovian SSE for the quadrature unraveling is
\begin{eqnarray}\label{QSSEOperatorForm} d_t\psizk{t}&=&\Big{[}
{-\frac{i}{\hbar}\hat{H}_{\rm int}(t)}-
(\hat{L}_{x}-\langle\hat{L}_{x}\rangle_t) \hat{Q}_{z}(t) \nl+
\Big{\langle}(\hat{L}_x-\langle\hat{L}_x\rangle_t) \hat Q_{z}(t)
\Big{\rangle}_t \nl+z(t)(\hat{L}-\langle\hat{L}\rangle_t)\Big{
]}\psizk{t}, \end{eqnarray} where
$\ket{{\psi}_{z}(t)}\equiv\ket{{\psi}_{\{X_{k}^{+}(t),Y_{k}^{-}(t)\}}(t)}$
and $\hat{L}_{x}=\hat{L}+\hat{L}\dg$. Again the operator
$\hat{Q}_{z}(t)$ is an ansatz to a functional derivative. In Ref.
\cite{GamWis02b} we outlined a perturbation technique for finding
this operator, if an exact solution cannot be found.

The differential equations for $X_{k}^{+}(t)$ and $Y_{k}^{-}(t)$
were shown, using the method of characteristics, to be
\begin{eqnarray} \label{X+Derivative}
{d_t}X_{k}^{+}(t)&=&g_k\cos(\Omega_{k} t)\langle \hat{L}_{x}\rangle_t ,\\
\label{Y-Derivative} {d_t}Y_{k}^{-}(t)&=&g_k\sin(\Omega_{k}
t)\langle \hat{L}_{x} \rangle_t.
\end{eqnarray}
Integrating these differential equation from time $0$ to $t$ we
get
\begin{eqnarray}
 X_{k}^{+}(t) &=&X_{k}^{+}(0)+\int_{0}^{t}g_k\cos(\Omega_{k} s)\langle
\hat{L}_{x}\rangle_s ds,\\
Y_{k}^{-}(t)&=&Y_{k}^{-}(0)+\int_{0}^{t}g_k\sin(\Omega_{k}
s)\langle \hat{L}_{x}\dg \rangle_s ds.
\end{eqnarray}
As in the position unraveling the random variables $X_{k}^{+}(0)$
and $Y_{k}^{-}(0)$ are chosen from the initial distribution. For
this unraveling and the initial condition
$\ket{\Psi(t)}=\ket{\{0_{k}\}}\ket{\psi(0)}$ this distribution is
\begin{equation}
{P}(\{X_{k}^{+},Y_{k}^{-}\},0)=\prod_{k>0}\frac{e^{-({X_{k}^{+}}^2+{Y_{k}^{-}}^2)}}{{\pi}}.
\end{equation}

To show that \erfs{X+Derivative}{Y-Derivative} can be derived from
the modal theory (velocity operator technique) we apply the
Hamiltonians defined in Sec. \ref{Sec3} to \erf{velocityO}. For
this unraveling the set of velocity operators $\{\hat{v}_{k}\}$
will be the union of $\{\hat{v}^{+}_{k}\}$ and
$\{\hat{v}^{-}_{k}\}$, where
\begin{eqnarray}
  \hat{v}^{+}_{k}(t)&=&\frac{-i}{\hbar}[\hat{X}^{+}_{k},\hat{H}_{\rm
  uni}]=g_{k}\hat{L}_{x}\cos(\Omega_{k}t),\\
  \hat{v}^{-}_{k}(t)&=&\frac{-i}{\hbar}[\hat{Y}^{-}_{k},\hat{H}_{\rm
  uni}]=g_{k}\hat{L}_{x}\sin(\Omega_{k}t),
\end{eqnarray} which are both real by definition. Substituting these velocity operators into
\erf{velocityField2} gives
\begin{eqnarray}
  {v}^{+}_{k}(\{X_{k}^{+},Y_{k}^{-}\},t)&=&g_{k}\bra{{\psi}_{\{X_{k}^{+},Y_{k}^{-}\}}(t)}\hat{L}_{x}
  \ket{{\psi}_{\{X_{k}^{+},Y_{k}^{-}\}}(t)}\nl\times
  \cos(\Omega_{k}t),\\
 {v}^{+}_{k}(\{X_{k}^{+},Y_{k}^{-}\},t)&=&g_{k}\bra{{\psi}_{\{X_{k}^{+},Y_{k}^{-}\}}(t)}
 \hat{L}_{x}\ket{{\psi}_{\{X_{k}^{+},Y_{k}^{-}\}}(t)}\nl\times
 \sin(\Omega_{k}t).
\end{eqnarray} Thus \erf{VelocityB} simply yields \erfs{X+Derivative}{Y-Derivative}.
Thus the modal theory gives the correct non-Markovian SSE.

\subsection{Coherent unraveling}

The last unraveling we consider is the coherent state unraveling.
This non-Markovian SSE was first presented by Di\'osi, Gisin, and
Strunz \cite{DioGisStr98}. In Ref. \cite{GamWis02} we showed that
it could be derived in the orthodox interpretation by considering
a bath measurement in terms of the Husimi POM \cite{Hus40}. This
has POM elements
\begin{equation}\label{POMexample}
  \hat{F}_{\{a_k\}}=\frac{1}{\pi^{K}}\ket{\{a_k\}}\bra{\{a_{k}\}},
\end{equation}
where $\hat{a}_{k}\ket{a_{k}}=a_{k}\ket{a_{k}}$ and $K$ is the
total number of modes. The noise function $z(t)$ for this
unraveling is
\begin{eqnarray}\label{Cohz}
  z(t)&=&\sum_{k}g_{k}a_{k}(t)e^{-i\Omega_{k}t}.
\end{eqnarray}
In Refs. \cite{DioGisStr98,GamWis02} it is shown that the
non-Markovian SSE for the coherent unraveling is
\begin{eqnarray}\label{CSSEOperatorForm} d_t\psizk{t}&=&\Big{[}
{-\frac{i}{\hbar}\hat{H}_{\rm int}(t)}-
(\hat{L}\dg-\langle\hat{L}\dg\rangle_t) \hat{C}_{z}(t) \nl+
\Big{\langle}(\hat{L}\dg-\langle\hat{L}\dg\rangle_t)
\hat{C}_{z}(t) \Big{\rangle}_t
\nl+z^{*}(t)(\hat{L}-\langle\hat{L}\rangle_t)\Big{ ]}\psizk{t},
\end{eqnarray} where
$\ket{{\psi}_{z}(t)}\equiv\ket{{\psi}_{\{a_{k}(t)\}}(t)}$. The
operator $\hat{C}_{z}(t)$ as in the previous unravelings is an
ansatz to a functional derivative. As in the quadrature unraveling
perturbation techniques exist for finding a perturbative solution
for this ansatz \cite{YuDioGisStr99,GamWis02b}.

Using the same procedure as the other unravelings, the
differential equation for $a_{k}(t)$ is
\begin{equation}\label{CoherentRandomDiff}
  d_{t}a_{k}^{*}(t)=g_{k}e^{-i\Omega_{k}t}\langle \hat{L}\dg\rangle_{t},
\end{equation} which integrates to give
\begin{equation}\label{CoherentRandomDVariable}
  a_{k}^{*}(t)=a_{k}^{*}(0)+\int_{0}^{t} g_{k}e^{-i\Omega_{k}s}\langle
\hat{L}\dg\rangle_s
  ds.
\end{equation} For an initial vacuum bath state, the random variable $a_{k}^{*}(0)$
is defined by the initial distribution
\begin{equation} \label{CProbPart}
{ P} (\{a_k\},0)= \prod_{k} \frac{{e^{-|a_{k}|^{2}}}}{\pi}.
\end{equation}

To show that the modal theory can be used to describe this
non-Markovian SSE we have to find the projector in ${\cal K}$
which is equivalent to the POM elements defined in
\erf{POMexample}. In Ref. \cite{GamWis03b} we showed that for a
single mode this projector is $\ket{x^{+},y^{-}}\bra{x^{+},y^{-}}$
where
 \begin{equation}\label{x+y-state}
\ket{x^{+},y^{-}}=\int
\frac{dx'}{\sqrt{2\pi}}\Ket{\frac{x^{+}-x'}{\sqrt{2}}}_{\rm
aux}\Ket{\frac{x^{+}+x'}{\sqrt{2}}}_{\rm uni}e^{iy^{-}x'},
\end{equation} where the states in the integrand are $x$ states. Thus the multimode projector used to define this
unraveling is
\begin{equation}
  \hat\Pi_{\{q_{k}\}}=\hat\Pi_{\{a_{k}\}} =\ket{\{x^{+}_{k},y^{-}_{k}\}}\bra{\{x^{+}_{k},y^{-}_{k}\}}_
  {\rm bath + aux}\otimes\hat{1}_{\rm sys},
\end{equation}
where $a_{k}$ is defined by
\begin{equation}
  a_{k}=x^{+}_{k}+iy^{-}_{k}.
\end{equation}
This allows us to define the operator $\hat{A}_{k}$, such that
\begin{equation}\label{}
\hat{A}_{k}\ket{x^{+},y^{-}}=a_{k}\ket{x^{+},y^{-}},
\end{equation}
as
\begin{equation}
  \hat{A}_{k}=\hat{x}^{+}_{k}+i\hat{y}^{-}_{k}=\hat a_{k}+\hat
  b\dg_{k},
\end{equation} which is a normal operator \cite{NieChu00}.
Here $\hat{x}^{+}_{k}$ and $\hat{y}^{-}_{k}$ are defined as
\begin{eqnarray}\label{x+y-etc}
\hat{x}^{+}_{k}&=&[\hat{a}_{k}+\hat{a}_{k}\dg+\hat{b}_{k}+\hat{b}_{k}\dg]/2, \\
\hat{y}^{-}_{k}&=&[-i\hat{a}_{k}+i\hat{a}_{k}\dg+i\hat{b}_{k}-i\hat{b}_{k}\dg]/2,
\end{eqnarray} where
$\hat{b}_{k}$ and $\hat{b}_{k}\dg$ are annihilation and creation
operators which act in ${\cal H}_{\rm aux}$. In this enlarged
Hilbert space the velocity operators are
\begin{eqnarray}
  \hat{v}^{+}_{k}(t)&=&-\frac{i}{\hbar}[\hat{x}^{+}_{k},\hat{H}_{\rm
  uni}\otimes\hat{1}_{\rm aux}]\nn \\
  &=&[g_{k}e^{i\Omega_{k}t}\hat{L}+g_{k}e^{-i\Omega_{k}t}
\hat{L}\dg]/{{2}},\\
  \hat{v}^{-}_{k}(t)&=&-\frac{i}{\hbar}[\hat{y}^{-}_{k},\hat{H}_{\rm
  uni}\otimes\hat{1}_{\rm aux}]\nn \\
  &=&[-ig_{k}e^{i\Omega_{k}t}\hat{L}+ig_{k}e^{-i\Omega_{k}t}
\hat{L}\dg]/{{2}}.
\end{eqnarray} With these velocity operators the
velocity fields become
\begin{eqnarray}
  {v}^{+}_{k}(\{x^{+}_{k},y^{-}_{k}\},t)&=&\bra{\psi_{\{x^{+}_{k},y^{-}_{k}\}}(t)}
  [g_{k}e^{i\Omega_{k}t}\hat{L}
 +g_{k}e^{-i\Omega_{k}t}\nl\times\hat{L}\dg] \ket{\psi_{\{x^{+}_{k},y^{-}_{k}\}}(t)}/2,\\
  {v}^{-}_{k}(\{x^{+}_{k},y^{-}_{k}\},t)&=&\bra{\psi_{\{x^{+}_{k},y^{-}_{k}\}}(t)}
  [-ig_{k}e^{i\Omega_{k}t}\hat{L}
 +ig_{k}\nl\times e^{-i\Omega_{k}t}\hat{L}\dg] \ket{\psi_{\{x^{+}_{k},y^{-}_{k}\}}(t)}/2.
\end{eqnarray}
Substituting these into \erf{VelocityB} gives
\begin{eqnarray}\label{SDEt}
  d_{t} x^{+}_{k}(t)&=&{[\langle\hat{L}\rangle_{t}g_{k}
  e^{i\Omega_{k}t} +\langle\hat{L}\dg\rangle_{t} g_{k} e^{-i\Omega_{k}
  t}]}/{{2}},\\
   d_{t} y^{-}_{k}(t)&=&{[-i\langle\hat{L}\rangle_{t}g_{k}
  e^{i\Omega_{k}t} +i\langle\hat{L}\dg\rangle_{t} g_{k} e^{-i\Omega_{k}
  t}]}/{{2}}.\nn \\
\end{eqnarray} Since $a_{k}=x^{+}_{k}+iy^{-}_{k}$ we once again easily obtain \erf{CoherentRandomDiff}
 as found from the orthodox theory.

\section{Discussion and Conclusions}\label{Sec5}

Given the success of continuous quantum measurement theory
\cite{Wis96b} in giving a nontrivial  interpretation of Markovian
SSEs, it is natural to seek a similar interpretation for diffusive
non-Markovian SSEs. It turns out that to give a meaning to such
non-Markovian SSEs we have to consider the modal interpretation of
quantum mechanics. This is because under the orthodox
interpretation of quantum mechanics, only the solution of the
non-Markovian SSE at a time $t$ can be given a meaning
\cite{GamWis02}. It corresponds to the state of the system at that
time given a measurement on the bath at that time yielding a
particular result. The bath cannot be measured continuously
because non-Markovian systems have a memory, so past measurements
of the bath will in general have a back-action which disrupts the
system's average evolution. Thus the solution at a particular time
may have an interpretation, but the linking of these solutions at
different times does not. The trajectory generated by the
non-Markovian SSE can be regarded only as a numerical tool for
calculating a conditioned state at a particular time.

However, under the modal interpretation, in particular a view
which is closest in line with Bell's beable theory
\cite{Bel84,Sud00}, we find that non-Markovian SSEs do have a
nontrivial interpretation. In this interpretation, the bath has
definite properties even if it is not measured, so the backaction
problem disappears. The system is treated as a purely quantum
system which, however, depends upon the values of the bath
properties: the values of the bath hidden variables. The evolution
of the system is generated by the non-Markovian SSE and the system
state guides the values of the hidden variables for the bath. The
bath hidden variables are similar to Bohmian hidden variables
\cite{Boh52}, as they obey a deterministic differential equation
with stochastic initial conditions. In fact, one unraveling we
have considered corresponds to Bohmian mechanics of the bath in an
interaction frame.

In this paper we also considered the quadrature and coherent
unravelings \cite{GamWis02}. The different unravelings are
determined by choosing which bath observable is to be given
property status. The noise $z(t)$ which appears in the
non-Markovian SSE is simply a linear combination of the
time-dependent values of the bath hidden variables. The quadrature
unraveling is defined such that $z(t)$ is real, while for the
coherent unraveling it is complex. In the Markovian limit the
former becomes the quantum trajectory for homodyne detection and
the latter becomes the quantum trajectory for heterodyne
detection. Thus quantum trajectories have both the standard
continuous measurement interpretation and the above modal
interpretation.

In conclusion it seems that to give diffusive non-Markovian SSE a
satisfying interpretation we must give up the orthodox
interpretation of quantum mechanics and consider the lesser known
but equally valid modal interpretation. In
Refs.~\cite{Lou01,BarLou02} it is claimed that the formalism
presented there does give a satisfying interpretation of
non-Markovian SSEs. If this is correct, we must conclude that said
formalism is a hidden-variable theory. The clarification of this
issue is beyond the scope of the current paper. Another problem
for future work would be to determine what choices for objective
properties of the bath will give rise to SSEs with a Markovian
limit. Finally, it would be interesting to use the modal theory to
develop discontinuous non-Markovian SSEs, such as those which
corresponds to a number state decomposition.

\acknowledgments

This work was supported by the Australian Research Council.


\begin{thebibliography}{}
\bibitem{Walter} W.T. Strunz (personal communication) has also sought an
interpretation of non-Markovain SSEs, but not along the lines
explored in this paper.
\end{thebibliography}
\end{document}